\documentclass[12pt,preprint]{aastex}
\shorttitle{Topology of HI Gas}
\shortauthors{Kim \& Park}

\begin{document}
%\twocolumn[
\title{Topology of HI Gas Distribution in the Large Magellanic Cloud}

\author{Sungeun Kim\altaffilmark{1} and Changbom Park\altaffilmark{2}}

\begin{abstract}
We have analyzed the H\,I aperture synthesis image of the Large Magellanic Cloud (LMC), using an 
objective and quantitative measure of topology to understand the H\,I 
distribution hosting a number of holes and clumps of various sizes in the medium.
The H\,I distribution shows different topology at four different chosen scales.
At the smallest scales explored ($19\sim 29$ pc), the H\,I mass is distributed
in such a way that numerous clumps are embedded on top of a low density
background. At the larger scales from 73 to 194 pc, it shows a generic
hole topology. These holes might have been formed mainly by stellar winds
from hot stars. At the scales from 240 to 340 pc, slightly above the disk 
scale-height of the gaseous disk, major clumps in the H\,I map change the distribution to 
have a slight clump topology.
These clumps include the giant cloud associations in the spiral arms and 
the thick filaments surrounding superholes. At the largest scales studied 
($390 \sim 485$ pc), the hole topology is present again. Responsible to the hole 
topology at this scale are a few superholes which seem mainly associated with 
supernova explosions in the outer disk. The gaps between the bar and spiral 
arms have a minor effect on the topology at this scale.
\end{abstract}


\altaffiltext{1}{Astronomy \& Space Science Department, Sejong University, 98
Kwangjin-gu,  Seoul 143-747, Korea; Email: sek@sejong.ac.kr}
%\altaffiltext{*}{Corresponding author: sek@sejong.ac.kr}
\altaffiltext{2}{Korea Institute for Advanced Study,
Dongdaemun-gu, Seoul 130-722, Korea}

\section{Introduction}

The topology of the interstellar medium (ISM) is critical in understanding the nature of 
the underlying physical structure that gives us the observed spatial and velocity structures. 
The global structure of the ISM is manifested by the
spatial and velocity structures of the neutral hydrogen gas. Atomic hydrogen is known to be 
an important component of the ISM (Burton 1992). Relatively high-resolution
H\,I images of nearby galaxies obtained with radio aperture synthesis interferometers have
shown overall clumpy H\,I distributions and various structures such as holes, shells, loops,
filaments, and bubbles (Brinks and Bajaja 1986; Deul and den Hartog 1990; Kamphuis, Sancisi,
and van der Hulst 1991; Puche et al. 1992; Staveley-Smith et al. 1997; Kim et al. 1998;
Stanimirovic et al. 1999; Thilker et al. 2000; Walter and Brinks 2001).
 %Issues are HI characteristics of (late-type) spiral galaxies. Namely, HI shells, bubbles,
%holes, blobs, and filaments.

One of the most striking results of these surveys is that the H\,I supergiant shells (SGSs)
occupy a large volume of the ISM as such the H\,I filaments seen in the Galaxy (Heiles
1984). About twenty SGSs with their diameters larger than 750 pc are identified in the Large
Magellanic Cloud (LMC), of which 1/3 are associated with optical counterparts detected in the
H$\alpha$ emission (Kim et al. 1999).  Many of these H\,I shells are found in regions of
very active star formation in the LMC. Massive stars interact with the ambient ISM through fast
stellar winds and supernovae ejecta to form interstellar shell structures with sizes ranging 
from 10 pc to greater than 1000 pc. However, the correlation between the H\,I shells and
122 OB stellar associations in the LMC (Lucke and Hodge 1970) is not very tight. No star
clusters were found at the center of the H\,I holes in Holmberg II (Rhode et al. 1999) and the
Galaxy (Heiles 1984). Deul and den Hartog (1990) have reported that the H\,I holes larger
than 500 pc are in general located at the interarm cavities and not likely to be produced by
the H\,II regions and OB associations. The energy required to produce such an H\,I hole is
at least 10$^{53}$ ergs (Kamphuis et al. 1991). These observations have risen an interesting
question about the origin of these structures and whether these H\,I holes have been formed
by the interaction between massive stars and the ISM or not. In fact, high-resolution
two-dimensional hydrodynamical simulations present that the large cavities of the ISM may be
formed by the nonlinear development of the combined thermal and gravitational instabilities,
without need for stellar energy injection in a galaxy modeling of the LMC (Wada et al. 2000).
In their study, dense clumps and filamentary structures are formed as a natural consequence
of the nonlinear evolution of the multiphase ISM. 
%A theory of turbulence leads to a log-normal 
%probability gas density function (Padoan \& Nordlund 2002).

In this paper, we present a method to disentangle the character of the structures seen in the
H\,I distribution in the LMC as a function of scale using the genus statistic. We find that the
genus curve can give information on the shape and topological properties of H\,I structure 
at different scales. The relationship between the amplitude of the genus curve and 
the slope of the power spectrum has been investigated in the current paper.

\section{Observational Data}
We use an H\,I aperture synthesis image from the high resolution HI survey of the LMC
with the Australia Telescope Compact Array (ATCA) at 1421 MHz with velocity coverage
from $-33$ to $627$ km/s (Kim et al. 1998). The results of H\,I aperture synthesis mosaic
of the LMC were made by combining data from 1344 separate pointing centers and correspond
to 11$^{\circ}$.1 $\times$ 12$^{\circ}$.4 on the sky. We have produced an HI brightness
temperature map of the LMC from the brightest H\,I component at each position. The spatial
resolution of the map is $50''\sim55''$ corresponding to 12$\sim$14 pc at the LMC's distance 
of 50.1 kpc (Alves 2004). At which, the LMC can be mapped with a high spatial resolution and 
both the interstellar structures and their underlying stars can be resolved and examined 
in great detail. Having low inclination angles, the LMC can be studied with little confusion 
along the line-of-sight. With small foreground and internal extinctions, the LMC can be 
easily observed at UV and X-ray wavelengths. The pixel size of the synthesis map is $20''$. 
Figure 1 shows the resulting HI
map of the LMC. The lines are circles of radii of 700, 600, and 500 pixels (from outside to
inward). We mainly choose to analyze the region within the circle of radius of 600 pixels
($200'$ or 3.0 kpc). This region contains most of the interesting structures and is not much
affected by the outer edge of the disk.
%\begin{figure}
%\includegraphics[scale=0.5]{fig1.eps}
%\caption{The H\,I brightness temperature map of the LMC used in the present
%work. Superposed are the circles of radii of 800 (outer most), 700, 600, and 500 pixels,
%where one pixel corresponds $20''$ or 5 pc at the distance of the LMC.
%}
%\label{fig1}
%\end{figure}

\section{The Genus Statistic}

We use the genus statistic to quantify the geometric shape of the projected
HI distribution of the LMC.  The intrinsic topology of iso-temperature or 
iso-density contours can be measured by the genus. 
The genus of an object has a meaning of the largest number of cuts 
that can be made through it without dividing it into two pieces. 
Following Melott et al. (1989) and Gott et al. (1990) we adopt,
for repeatedly or infinite connected contours in the plane, 
a modified definition of the two-dimensional genus
\begin{equation}
G(\nu) = \mbox{\# of isolated high density regions} - 
         \mbox{\# of low density holes}.
\end{equation}
This genus equals to 1 minus the mathematically-defined genus.
For example, the genus of a ring will be zero because it has one connected 
high density region and one hole.  On the other hand, the genus of a disk 
is plus one because it has one isolated high density region and no hole.
In the case of the HI map of the LMC
we first smooth the map over a scale we are interested in, 
and construct the iso-density contours at a set of threshold levels.
The genus is then calculated by integrating the local curvature along the
contours. With a threshold level $\nu$ the mathematical form of the genus is
\begin{equation}
G(\nu) = {1\over {2\pi}} \sum_{i} \oint_{C_i} \kappa ds,
\end{equation}
where $\kappa$ is the local (signed) curvature on an isodensity contour $C_i$. 
We define the sign of $\kappa$ so that it is positive when a contour 
encloses a high density region counterclockwise (Gott et al. 1990;
Park et al. 1992; Park, Gott, \& Choi 2001).
A genus curve is obtained by measuring the genus at a set of threshold levels. 
A good feature of the genus statistic is that the analytic formula of 
the genus curve is known for a Gaussian random field. 
The genus for a Gaussian random phase field is
\begin{equation}
g(\nu) = A \nu  e^{-\nu^2/2},
\end{equation}
where the amplitude $A$ depends only on the shape of the power spectrum of
the field but not on its amplitude (see eqs. 8 and 9).
Any deviation of a measured genus curve from this formula
is evidence for non-Gaussianity, and reveals the nature of the geometrical
shape of the fluctuation.
The threshold levels are chosen so that they can represent area fractions.
A threshold level with a label $\nu$ corresponds to the area fraction
\begin{equation}
f_{\nu} = {1\over(2\pi)^{1/2}}\int_{\nu}^{\infty} e^{-x^2/2} dx.
\end{equation}
Contours with $\nu=1.0, 0.0$, and $-1.0$ enclose 16, 50, 
and 84\% of the total area, respectively.

To get an idea of what the genus curve indicates about the shape of structures 
in the disk of a galaxy, we show in Figure 2 the genus curves of three toy models. 
The solid line is the genus curve of a disk filled with randomly fluctuating 
but statistically uniform matter smoothed over $\lambda_{\rm FWHM}=2'$
by a Gaussian filter. This Poission matter fluctuation yields a Gaussian 
random field, and the genus curve has the form of equation (3).
The dashed line with filled circles is for a uniform disk with additional 857 
randomly distributed clumps of diameter of $4'$.
In this case of clump topology, the genus curve is shifted to the left, 
and becomes asymmetric.
The dotted line with open circles is for a uniform disk with 857 randomly
distributed empty holes of diameter of $4'$.  The genus curve is now shifted 
to the right and also becomes asymmetric, indicating that the shape of 
the low density regions is altered by the holes.
Since the area fraction is fixed at a given $\nu$, the low amplitude of
the genus curve means fewer number of structures with bigger size
compared to the expectation for a Gaussian field.
Therefore, the genus curve tells not only the deviation of a field from
the Gaussian one, but also the statistical nature of the field too.
Even though the true nature of the HI distribution in the LMC is 
three dimensional, the projected view of the distribution shown in Figure 1
reveals a wealth of interesting structures, which deserve a two dimensional
topology study. The two dimensional genus statistic has been applied 
to the distribution of galaxies when the data is the projected distribution 
of galaxies on the sky or the distribution of galaxies in a thin slice 
(Melott et al. 1989; Park et al. 1992).

%\begin{figure}
%\includegraphics[scale=0.4]{fig2.eps}
%\caption{Genus curves of three toy models of a circular disk. The solid line
%is the genus curve of a disk with random and uniform matter distribution.
%The dashed line with filled circles is for a uniform disk with additional
%857 clumps of diameter of $4'$, and the dotted line with open circles
%is for a uniform disk with 857 holes of diameter of $4'$.
%The size of the disk is set to 200 arc minutes, and the Gaussian smoothing
%length is $2'$.}
%\label{fig2}
%\end{figure}

\section{Results}

%\begin{figure}
%\includegraphics[scale=0.5]{fig3.eps}
%\caption{The HI maps of the LMC smoothed over * (left) and ** (right) FWHM
%by a Gaussian filter, catching structures at different scales.
%}
%\label{fig3}
%\end{figure}

To measure the genus we smooth the HI image of the LMC and cut the circular
area of the disk within $200'$ from the center.
%Two examples are shown
%in Figure 3, where the Gaussian smoothings of * and ** FWHM are applied.
%A wealth of interesting objects are seen in both maps with emphasis of
%structures at different scales.
The genus curves of the HI distribution of the LMC for 12 selected
smoothing scales are shown in Figure 3.
In each panel three genus curves are plotted and their
Gaussian smoothing lengths are denoted in units of arcminutes.
The solid line in each panel is the Gaussian curve (eq. 3)
best fit to the filled circles
corresponding to the median smoothing length.
Open circles represent the shortest smoothing length.
The genus curve for $\lambda_{\rm FWHM}=1.7'$ (open circles
in the lower right panel) smoothing shows no shift, but the negative side has
an amplitude higher than the positive side.
%\begin{figure}
%\includegraphics[scale=0.45]{fig3.eps}
%\caption{The genus curves of the HI map of the LMC at various smoothing
%scales. The smoothing lengths are in units of pixels and the radius of
%the circular region under study is set to $200'$. In each panel
%open circles are for the shortest smoothing length, and squares are for
%the longest smoothing length. The solid line is the Gaussian curve
%best fitting to the filled circles
%corresponding to the median smoothing length.
%}
%\label{fig3}
%\end{figure}
This means that the low density regions are broken into
numerous pieces while the high density regions are relatively more
connected into fewer number of pieces at the same volume fraction.
Therefore, the genus curve indicates that there are coherent high density 
clumps distributed on a noisy (low fluctuation amplitude) background 
at this shortest scale.
On the other hand, the genus curves corresponding to $\lambda_{\rm FWHM}
=10'$ or $13'$ (filled circles or squares in the lower left panel) shows
a slight shift to the right and the low density part has a lower
amplitude. This is a generic shape of a hole topology as shown in Figure 2.
At larger smoothing lengths the genus curves become noisier and
the topological structure of the HI disk is not obvious from Figure 3.

To make more quantitative measurements of the shift and asymmetry of 
the genus curve, we develop pseudo-statistics derived from the genus curve.
We first find the best-fit amplitude of the Gaussian genus curve for
each measured genus curve over the interval $-2\leq \nu \leq +2$.
We quantify the shift by
\begin{equation}
\Delta\nu = \int_{-1}^{1} \nu G(\nu)d\nu /\int_{-1}^{1} G_{G}(\nu)d\nu,
\end{equation}
where $G_{G}$ is the best-fit Gaussian genus curve (Park et al. 2001).
The asymmetry of the genus curve at high density levels is quantified by
\begin{equation}
A_C = \int_{0.7}^{2} G(\nu)d\nu /\int_{0.7}^{2} G_{G}(\nu)d\nu.
\end{equation}
If the measured genus is lower than the best-fit Gaussian curve
at high threshold levels, the parameter $A_C$ will be less than 1,
meaning that the high density regions are more connected into fewer larger
pieces compared to the Gaussian case.
The asymmetry parameter $A_H$ at low density levels is similarly defined
with the integration interval from $-0.7$ to $-2.0$,
\begin{equation}
A_H = \int_{-2}^{-0.7} G(\nu)d\nu /\int_{-2}^{-0.7} G_{G}(\nu)d\nu.
\end{equation}
If the low density regions are well-connected to one another
or dominated by a few large holes,
the genus becomes lower than Gaussian expectation and the parameter
$A_H$ becomes less than 1.

Figure 4 shows interesting variations of the shift and asymmetry parameters
as functions of smoothing scale. The circles and triangles in the
upper panel are the shift parameters measured from the circular areas
with radii of $200'$ and $167'$, respectively.
The difference between the two cases is a measure of the uncertainty
in the shift parameter.
They show that the results are generally robust against the choice
of the outer boundary. In the bottom panel
the asymmetry parameters $A_{C}$ (filled circles)
and $A_{H}$ are plotted. These parameters
reveal an interesting change of topology as the scale changes.
At the smallest scales studied ($1.3'\sim 2'$ or $19\sim 29$ pc), the shift parameter is nearly
zero while $A_C <1$ and $A_H>1$, which was seen in the lower right panel
of Figure 3. This corresponds to the situation when there are clumps
embedded in the background HI distribution which has low-amplitude random fluctuations. 
At high density thresholds the clumps
are detected and their number density falls below that of Gaussian random
field. But at low thresholds the isolated low density holes are found
in the fluctuating background, and their number density exceeds that
of the best-fit Gaussian model.

At the next larger scales ($5'\sim 13'$ or $73\sim 194$ pc) the HI map 
shows a generic hole topology. The shift parameter is positive,
and the genus curve is asymmetric in the sense that the low density holes
are fewer (or larger in size) than the Gaussian expectation. This was also
noticed in the lower left panel of Figure 3.
The transition from the clump to the hole topology occurs at about $3'$ or 48 pc.
At the still larger scales of $17'\sim 23'$ or $240\sim 340$ pc the asymmetry 
of the genus curve remains the same, but the shift parameter suddenly drops 
to slightly negative values. The clump topology is not statistically significant 
when the uncertainty at these scales is taken into account, but the change 
in topology is clear. It seems that there are large dense clumps as well as 
large holes of similar sizes at these scales. The large clumps are large 
enough to show up in the shift parameter, but are not prominent and numerous 
enough to change the whole topology of the disk. The scale is about 1/10 
of the size of disk, and the corresponding clumps are mainly distributed 
along the bar and spirals arms. Some of them are filaments around the biggest holes.

At the largest scales explored ($27'\sim 33'$ or $390\sim 485$ pc) the genus curves 
are very noisy because of the small number of resolution elements at these large smoothing scales.
However, the genus-related parameters suggest a hole topology at these scales. 
The shift parameter, though uncertain, is positive, and the genus curve is 
asymmetric again with low amplitude at low density thresholds as in the second case above.
There are a few big holes causing this behavior of the genus curve. 
They are mainly the roughly circular giant holes in the upper left part of the disk 
rather than the gaps between spiral arms (see Figure 1). The lower right part, 
where there are several connected holes of smaller sizes located between the spiral arms, 
also contributes to the hole shift.

%\begin{figure}
%\includegraphics[scale=0.5]{fig4.eps}
%\caption{Variation of the shift and asymmetry parameters
%as a function of smoothing length. The circles and triangles in the
%upper panel are the shift parameters measured from the circular area
%within radii of $200'$ (or 600 pixels) and $167'$, respectively.
%In the lower panel the asymmetry parameters $A_C$ (filled circles)
%and $A_H$ (open circles) are plotted.
%}
%\label{fig4}
%\end{figure}

%In addition to topology
%we now study the power spectrum of the HI fluctuation
%because the characteristic scales showing changes of
%topology might be associated with the shape of the power spectrum.
In addition to the topology of the H\,I distribution
we also study its power spectrum because the characteristic scales 
showing topological changes could appear in the shape of power spectrum as well. 
The amplitude of the genus curve can be used to explore the slope of
the power spectrum near the smoothing scale. Consider a two dimensional
Gaussian random field, whose power spectrum is $P(k)$, which is smoothed
by a Gaussian kernel $F(k) = {\rm exp}(-k^2 R_G^2 /2)$ over a smoothing length
$R_G$. Then the genus curve per unit area in equation (2) is given by
\begin{equation}
A = {1\over{(2\pi)^{3/2}}} {{\langle k^2\rangle}\over 2},
\end{equation}
where
\begin{equation}
\langle k^2\rangle = \int k^2 P(k)F^2(k) dk /\int P(k) F^2(k) dk.
\end{equation}
When the power spectrum is a power-law one, $P(k)\propto k^n$,
the moment of power spectrum is given by
\begin{equation}
\langle k^2\rangle = (n_{\rm eff}+2)/2 R_G^2.
\end{equation}
Note that the Gaussian smoothing length is related with the FWHM by
$\lambda_{\rm FWHM} = 2\sqrt{2 {\rm ln 2}} R_G$.
For Poission noise $n=0$, and the total genus at $\nu=1$ is expected to be
3350 for an area with radius of $200'$ and a smoothing scale
of $\lambda_{\rm FWHM}=2'$. This is exactly what is seen for the random
mass distribution in Figure 2 (solid curve).

%\begin{figure}
%\includegraphics[scale=0.5]{fig5.eps}
%\caption{(upper) The genus amplitude $A$ per unit square degree as a function
%of smoothing scale. (lower) The effective power index of the power
%spectrum of the HI matter fluctuation calculated from the genus density
%of the upper panel. The filled circles are for the area of radius of $200'$,
%and the open squares are for $167'$.
%}
%\label{fig5}
%\end{figure}

The upper panel of Figure 5 shows the amplitude of the genus curve
per unit square degree as a function of the
smoothing scale. For a Gaussian matter fluctuation with a perfect power-law power
spectrum it will scale as $\lambda^{-2}$. But it is dropping faster
than that at larger scales.  The bottom
panel shows the effective power index of the power spectrum at
the smoothing scale calculated from equation (10) assuming that the HI
distribution is a Gaussian random field. 
Even though the HI distribution is not exactly Gaussian, the deviation is 
not great as can be seen in Figures 3 and 4 and the index of the power 
spectrum measured by eq. (10) can serve as an estimate of the true index. 
Figure 5 shows that the power index continuously
increases from the smallest scale studied to about $\lambda_{\rm FWHM}=20'$, 
where it stays more or less constant at $-1$.
%We do not see a clear relation between the features in the power spectrum
%and those in the topology except that
%the $20'$ scale is where the clumps and holes are competing to each other.
The $20'$ scale is where the clump and hole topologies are competing with each other.
%We only see a relation between the features of the topological structure 
%and the power spectrum at the scales 20$? scale where the clumps and holes 
%are competing with each other. 
The power index of $-1.8$ at the smallest scale means that the HI mass
fluctuation is nearly scale-free. On the other hand, $n\gg -2$ at
larger scales means that the mass fluctuation is higher at small scales
and lower at large scales.
The constancy of the power index at scales larger than $20'$ might mean
there is some mechanism generating additional matter fluctuations at those scales and above.

\section{Discussion}

We performed a topological analysis of the H\,I map of the LMC 
which hosts a number of holes and clumps of various sizes. 
%We have analyzed the H\,I map of the LMC using an objective and quantitative
%measure of topology to understand the H\,I distribution of the LMC hosting a number
%of holes and clumps of various sizes.
The H\,I distribution shows different topology at four different scales.
At the smallest scale explored ($19\sim 29$ pc) the H\,I mass is distributed
in such a way that numerous clumps are embedded on top of a low density
background. At the larger scales from 73 to 194 pc it shows a generic
hole topology. These holes might have been mainly formed by stellar wind 
from massive stars and SNe.  Therefore, the structure of the neutral atomic 
interstellar gas is dominated by numerous small clumps and relatively larger holes 
at the scales less than the scale-height of the gaseous disk, which has been
estimated to be approximatedly 180 pc (Kim et al. 1999).

At the scales from 240 to 340 pc major clumps in the H\,I map change the distribution to
a slight clump topology.  These H\,I clumps include the giant cloud associations 
in the spiral arms and the thick filaments surrounding large holes. 
At the largest scales studied ($390 \sim 485$ pc) the hole topology is again detected. 
Responsible to the hole topology are a few large holes which seem mainly associated with 
supernova explosions at outer disk. The gaps between the bar and spiral arms have
minor effects.

We have also measured the effective power index of the power spectrum of the H\,I 
map. The H\,I distribution has a nearly scale-free power spectrum at the smallest 
scales explored.  But the power index continuously increases at larger scales. 
This means that the H\,I mass fluctuation at these scales is 
dominated by the power at smaller scales.
At scales larger than 290 pc the power index stays at a roughly constant value of $-1.05\pm 0.05$. 
This transition scale is comparable to the H\,I scale-height estimated from 
the average vertical velocity dispersion and the average surface density of 
both the H\,I and the stellar components of the LMC disk (Kim et al. 1999).
However, a wavelength of 290 pc is still shorter than the warm H\,I diffuse layer
having a thickness of 350 pc estimated from the basis of supershell model for LMC 2 
(Wang \& Helfand 1991).

\acknowledgments
We thank Yun-Young Choi for her help for illustration.
SK and CBP acknowledge the support of the Korea Science and Engineering Foundation (KOSEF) through the Astrophysical Research Center for the Structure and Evolution of the Cosmos (ARCSEC). 

{}

\begin{figure}
%\plotone{fig1.eps}
\caption{The H\,I brightness temperature map of the LMC (Kim et al. 1998) used in 
the present work. Superposed are the circles of radii of 700 (outer most), 
600, and 500 pixels, where one pixel corresponds $20''$ or 5 pc at the distance of 
the LMC. }
\label{fig1}
\end{figure}

\begin{figure}
\plotone{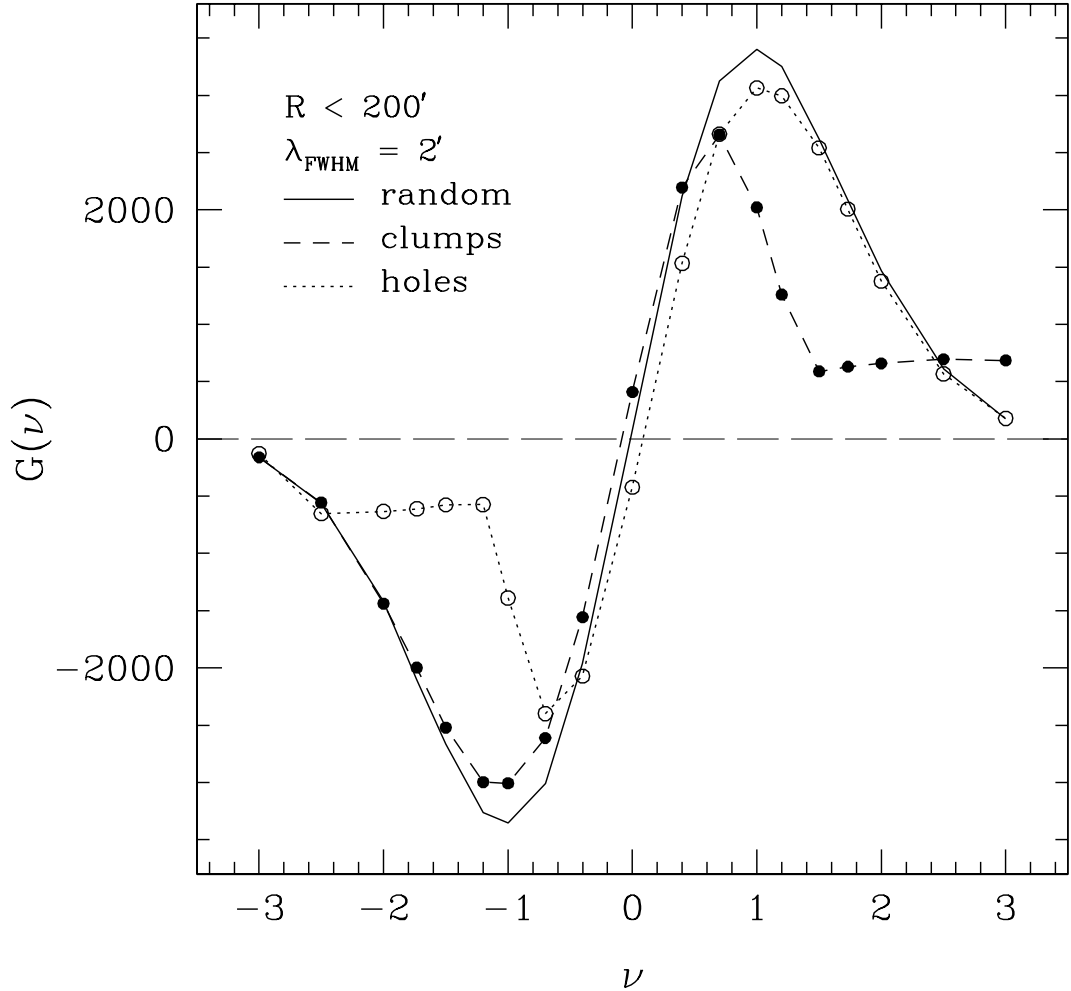}
\caption{Genus curves of three toy models of matter distribution in a circular disk. 
The solid line
is the genus curve of a disk with Poission matter distribution.
The dashed line with filled circles is for a disk of Poission matter distribution 
with additional 857 clumps of diameter of $4'$, and the dotted line with open circles
is for a disk with 857 holes of diameter of $4'$ in a Poission matter distribution.
The size of the disk is set to 200 arc minutes, and the Gaussian smoothing
length is $2'$.}
\label{fig2}
\end{figure}

\begin{figure}
\plotone{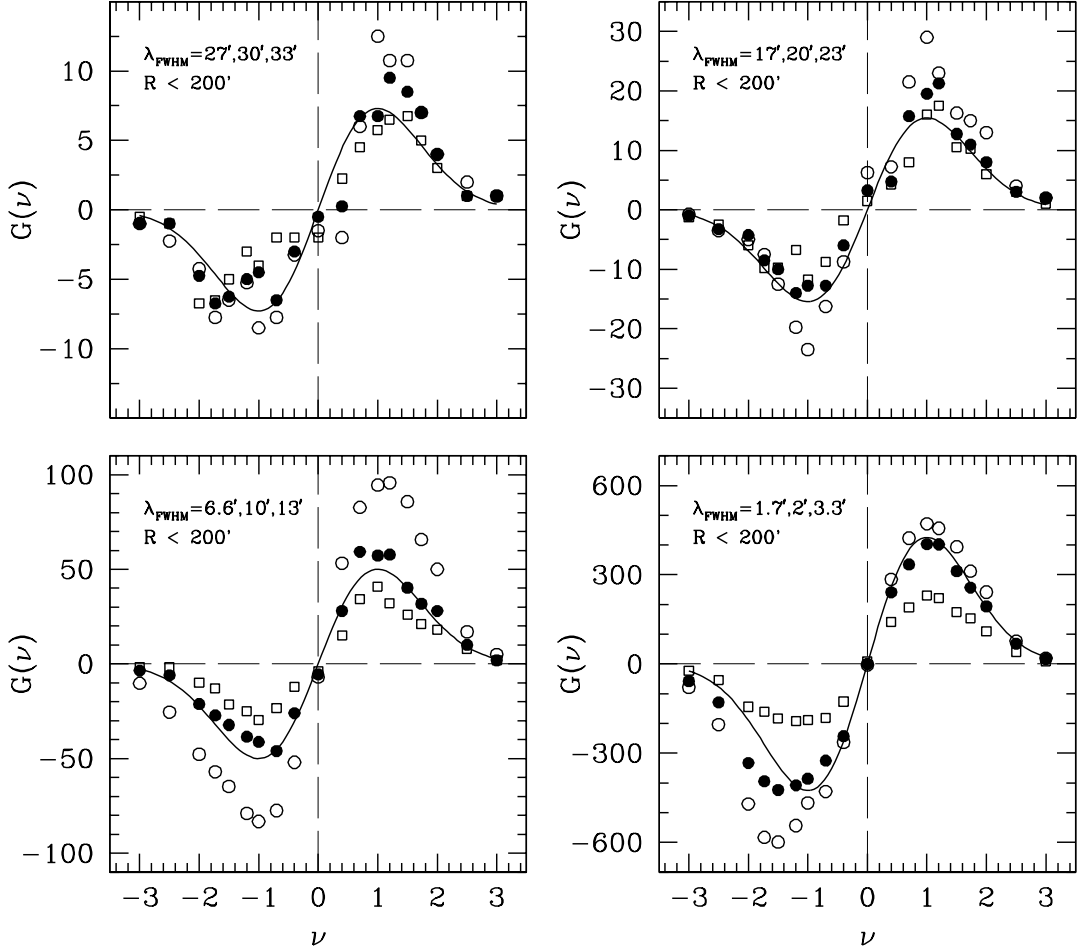} \caption{The genus curves of the HI map of the
LMC at various smoothing scales. The smoothing lengths are in
units of pixels and the radius of the circular region under study
is set to $200'$. In each panel open circles are for the shortest
smoothing length, and squares are for the longest smoothing
length. The solid line is the Gaussian curve best fitting to the
filled circles corresponding to the median smoothing length. }
\label{fig3}
\end{figure}

\begin{figure}
\plotone{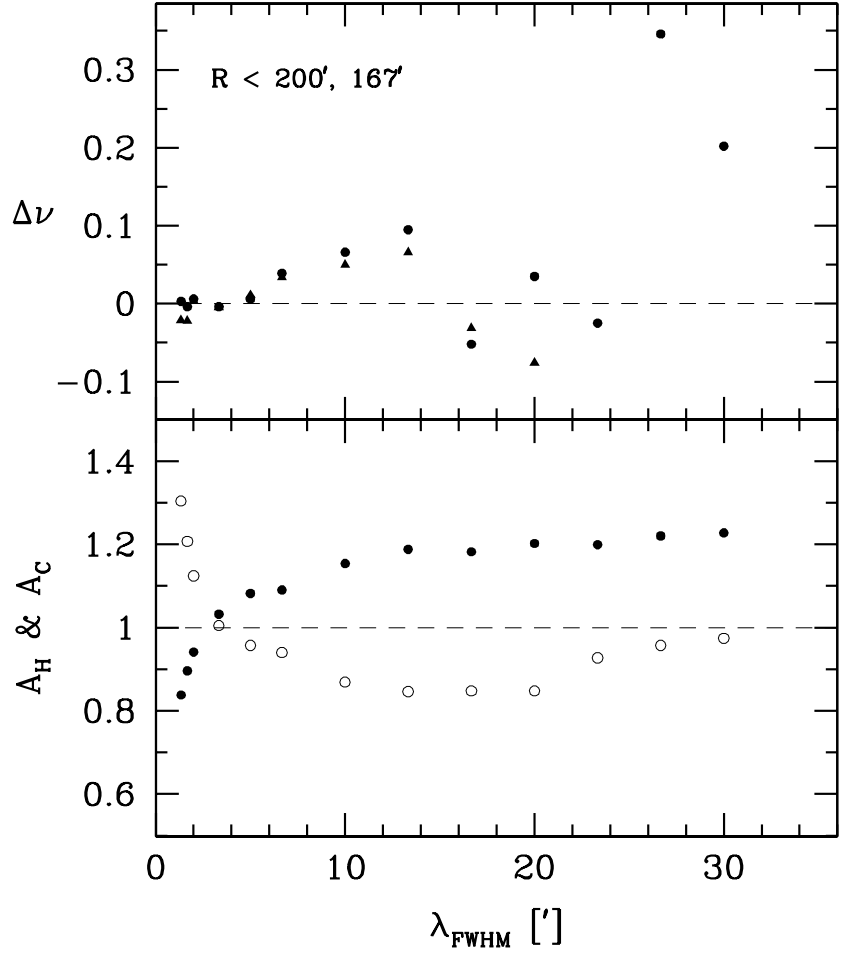} \caption{Variation of the shift and asymmetry
parameters as a function of smoothing length. The circles and
triangles in the upper panel are the shift parameters measured
from the circular area within radii of $200'$ (or 600 pixels) and
$167'$, respectively. In the lower panel the asymmetry parameters
$A_C$ (filled circles) and $A_H$ (open circles) are plotted. }
\label{fig4}
\end{figure}

\begin{figure}
\plotone{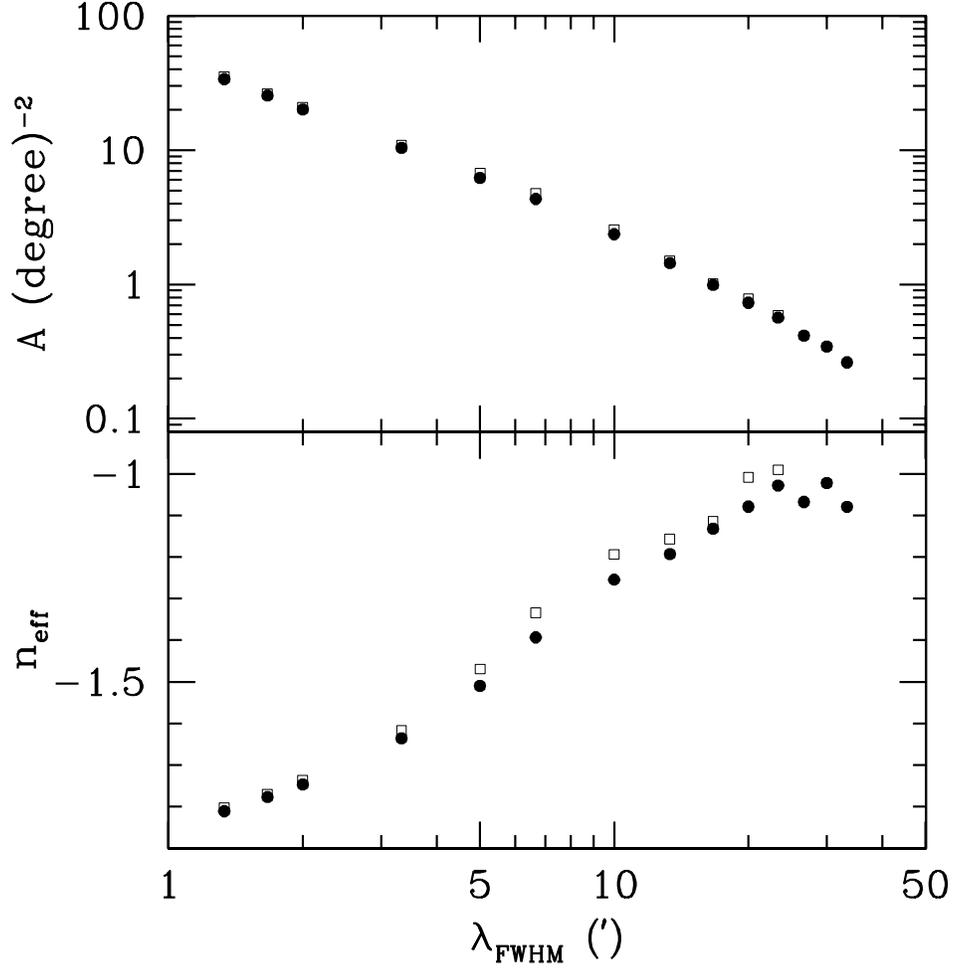} \caption{(upper) The genus amplitude $A$ per
unit square degree as a function of smoothing scale. (lower) The
effective power index of the power spectrum of the HI matter
fluctuation calculated from the genus density of the upper panel.
The filled circles are for the area of radius of $200'$, and the
open squares are for $167'$. } \label{fig5}
\end{figure}
\end{document}